\documentclass[twocolumn]{aastex63}

\usepackage{lipsum}
\usepackage{kotex}
\usepackage{xcolor}
\usepackage{enumerate}
\usepackage{multirow}
\usepackage{graphicx}
\usepackage[normalem]{ulem}
\usepackage{hyperref}
\usepackage{tabularx}
\usepackage{verbatim}
\usepackage{booktabs}
\usepackage{afterpage}
\usepackage{amsmath}
\usepackage{lineno}
\usepackage{hyperref}
\hypersetup{
    colorlinks=true,
    linkcolor=blue,
    filecolor=magenta,      
    urlcolor=blue,
}

\received{\today}
\revised{}
\accepted{}
\submitjournal{ApJ}

\shorttitle{U-type Warped Disks}
\shortauthors{Zee et al.}
\graphicspath{{./}{figures/}}

\begin{document}

\title{Warped Disk Galaxies. I. Linking U-type Warps in Groups/Clusters to Jellyfish Galaxies}

\correspondingauthor{Suk-Jin Yoon}
\email{sjyoon0691@yonsei.ac.kr}

\author[0000-0003-0960-687X]{Woong-Bae G. Zee$^*$}
\affiliation{Department of Astronomy, Yonsei University, Seoul, 03722, Republic of Korea}
\affiliation{Center for Galaxy Evolution Research, Yonsei University, Seoul, 03722, Republic of Korea}

\author[0000-0002-1842-4325]{Suk-Jin Yoon$^*$}
\affiliation{Department of Astronomy, Yonsei University, Seoul, 03722, Republic of Korea}
\affiliation{Center for Galaxy Evolution Research, Yonsei University, Seoul, 03722, Republic of Korea}

\author[0000-0001-7075-4156]{Jun-Sung Moon}
\affiliation{Department of Astronomy, Yonsei University, Seoul, 03722, Republic of Korea}
\affiliation{Center for Galaxy Evolution Research, Yonsei University, Seoul, 03722, Republic of Korea}

\author[0000-0003-3791-0860]{Sung-Ho An}
\affiliation{Department of Astronomy, Yonsei University, Seoul, 03722, Republic of Korea}
\affiliation{Center for Galaxy Evolution Research, Yonsei University, Seoul, 03722, Republic of Korea}

\author[0000-0003-2922-6866]{Sanjaya Paudel}
\affiliation{Department of Astronomy, Yonsei University, Seoul, 03722, Republic of Korea}
\affiliation{Center for Galaxy Evolution Research, Yonsei University, Seoul, 03722, Republic of Korea}

\author{Kiyun Yun}
\affiliation{Max-Planck-Institut für Astronomie, Königstuhl 17, D-69117 Heidelberg, Germany}

\def\thefootnote{*}\footnotetext{These authors contributed equally to this work.}\def\thefootnote{\arabic{footnote}}

\begin{abstract}
Warped disk galaxies are classified into two morphologies: S- and U-types.
Conventional theories routinely attribute both types to galactic tidal interaction and/or gas accretion, but reproducing of U-types in simulations is extremely challenging.
Here we investigate whether both types are governed by the same mechanisms using the most extensive sample of $\sim$8000 nearby (0.02\,$<$\,z\,$<$\,0.06) massive ($M_{*}/M_{\odot}$\,$>$\,$10^9$) edge-on disks from SDSS.
We find that U-types show on average bluer optical colors and higher specific star formation rate (sSFR) than S-types, with more strongly warped U-types having higher sSFR.
We also find that while the S-type warp properties correlate with the tidal force by the nearest neighbor regardless of the environment, there is no such correlation for U-types in groups/clusters, suggesting a non-tidal environmental could be at play for U-types, such as ram pressure stripping (RPS).
Indeed, U-types are more common in groups/clusters than in fields and they have stellar mass, gas fraction, sSFR enhancement and phase-space distribution closely analogous to RPS-induced jellyfish galaxies in clusters.
We furthermore show that the stellar disks of most RPS galaxies in the IllustirsTNG simulation are warped in U-shape and bent in opposite direction of stripped gas tails, satisfying theoretical expectations for stellar warps embeded in jellyfishes.
We therefore suggest that despite the majority of U-types that live in fields being still less explained, RPS can be an alternative origin for those in groups/clusters.

\end{abstract}

\section{Introduction} \label{sec:intro}

Observations over the past few decades showed that the warped disk structure is common in the local universe. 
More than half of nearby edge-on disk galaxies observed in optical and radio passbands exhibit warps at the outskirts of disks (e.g., \citealt{1990MNRAS.246..458S}; \citealt{1991wdir.conf..181B}; \citealt{1998A&A...337....9R}; \citealt{2002A&A...382..513R}; \citealt{2002A&A...391..519C}; \citealt{2002A&A...394..769G}; \citealt{2003A&A...399..457S}; \citealt{2006NewA...11..293A}; \citealt{2016MNRAS.461.4233R};  see also \citealt{2019NatAs...3..320C}; \citealt{2020ApJ...905...49C}; \citealt{2021ApJ...912..130C} for the Milky Way's warp).
Optical stellar warps are, in general, weaker than HI gaseous warps; however, the incidence of optical warps is as prevalent as HI warps (e.g., \citealt{1990ApJ...352...15B}; \citealt{2002A&A...391..519C}; \citealt{2002A&A...394..769G}; \citealt{2010A&A...519A..53G}; \citealt{2008MNRAS.389...63C}). 
The morphology of warped disks is classified into two types, S- (integral-shaped) and U-type (bow-shaped) (\citealt{1998A&A...337....9R}; \citealt{2006NewA...11..293A}).

Galactic warps are often taken as results of galaxy--galaxy interactions.
For instance, simulations by \citet{2018MNRAS.481..286L} and \citet{2018Natur.561..360A} reproduced the grand-design S-shaped warp of the Milky Way using the orbiting Sagittarius dwarf and Magellanic Clouds. 
\citet{2014ApJ...789...90K} and \citet{2017MNRAS.465.3446G} suggested that the fly-by encounter is another warp formation mechanism. 
\citet{2020MNRAS.498.3535S} used IllustrisTNG simulation to investigate the origin of warped galaxies. 
The authors focused only on S-type warps and showed that $\sim$30\% of S-types are constructed by tidal interactions with $\sim$15\% being minor mergers and $\sim$15\% fly-by encounters. 
Observationally, some studies suggested that interacting galaxies are more frequently warped than non-interacting galaxies (\citealt{1998A&A...337....9R}; \citealt{2001A&A...373..402S}; \citealt{2006NewA...11..293A}). 
There is, however, only a weak correlation (\citealt{1990A&A...233..333K}; \citealt{1998A&A...337....9R}; \citealt{2001A&A...373..402S}) or even anti-correlation (\citealt{2002A&A...394..769G}) between the frequency of warps and the local environment, at variance with the conventional tidal scenario.
Moreover, several isolated galaxies in the field exhibit warped disks (\citealt{1990ApJ...352...15B}; \citealt{1997A&A...321..754V}; \citealt{2007JKAS...40....9A}; \citealt{2008A&A...488..511L}). 
Thus, other alternative warp formation mechanisms have been suggested, including cold gas flow (\citealt{1989MNRAS.237..785O}; \citealt{1999MNRAS.303L...7J}; \citealt{2010MNRAS.408..783R}; \citealt{2018MNRAS.474..254R}), interaction with the intergalactic accretion onto disks (\citealt{2002A&A...386..169L}; \citealt{2006MNRAS.365..555S}), ram-pressure by galaxies' movement with respect to the inter-galactic medium (\citealt{2014MNRAS.440L..21H}), misalignment between stellar disks and prolate/oblate dark matter halos (\citealt{1988MNRAS.234..873S}; \citealt{1991ApJ...376..467K}; \citealt{1995MNRAS.275..897N}; \citealt{2000MNRAS.311..733I}; \citealt{2009ApJ...696.1899J}; \citealt{2009ApJ...703.2068D}; \citealt{2021arXiv211013964S}), and interaction with the intergalactic magnetic field (\citealt{1990A&A...236....1B}; \citealt{1998A&A...332..809B}).
However, which mechanism dominates the warp formation remains debated.

When it comes to U-type warps, many observations witnessed conspicuously warped U-types (\citealt{2002A&A...382..513R}; \citealt{2003A&A...399..457S}; \citealt{2006NewA...11..293A}), but their physical origin is still a puzzle.
Most of conventional studies simply assumed that S- and U-types are results of the same warp formation mechanism, such as the tidal interaction. 
However, simulations of tidal interactions preferentially create S-types with no or few U-types (e.g., \citealt{1995ApJ...455L..31W}; \citealt{2000ApJ...534..598V}; \citealt{2006ApJ...641L..33W}; \citealt{2008MNRAS.388..697M}; \citealt{2013MNRAS.429..159G}; \citealt{2014ApJ...789...90K}; \citealt{2017MNRAS.465.3446G}; \citealt{2018MNRAS.481..286L}; \citealt{2013MNRAS.429..159G}; \citealt{2018Natur.561..360A}). 
On the other hand, \citet{2008ASPC..390..359L} elaborated that intergalactic gas accretion can produce both types, in that S- and U-types are respectively made via angular and linear momentum transmission during gas accretion. When intergalactic gas is accreted from the nearly vertical direction to the galactic disk, galaxies are bent into the U-shaped morphology. 
Alternatively, the simulation by \citet{2020MNRAS.498.1080K} showed that the warped disk’s morphology depends on infalling galaxies’ relative direction to the dark matter particles. 
When a galaxy moves through dark matter halo, tidal friction induced by the over-density region behind the drifting galaxy produce U-shaped disks.

Our main questions in this study are: ($a$) whether S- and U-type warps are crafted by the same mechanism? and ($b$) how can we explain the existence of U-type warps?
To address them, we construct the most extensive catalog of nearby edge-on warped disk galaxies from Sloan Digital Sky Survey (SDSS) Data Release 7 (DR7). 
We for the first time distinguish the intrinsic characteristics of S- and U-type warped galaxies and report the discovery of their discrepancy.
We propose a new possibility that U-type warps in groups/clusters could be related to ram-pressure stripping (RPS). 
The paper is organised as follows. 
In Section~\ref{sec:maths}, we introduce our data and explain how we measure the warp structure using our newly developed, automated scheme. 
In Section~\ref{sec:result}, we show the discrepancy in several intrinsic properties between S- and U-types, including the optical color, specific star formation rate (sSFR) and environmental effects.
In Section~\ref{sec:kine}, we compare the morphologies, kinematics within groups/clusters, sSFR, stellar masses and gas fraction of S- and U-types with those of RPS galaxies and discuss the possible jellyfish origin of U-type warped disks.
In Section~\ref{sec:conc} we summarise our results.

\section{Data and Methodology} 
\label{sec:maths}

\subsection{Observational Data}

We select sample galaxies from the SDSS DR7 Legacy Survey (\citealt{2009ApJS..182..543A}). The SDSS Legacy Survey is a large optical imaging survey that provides a map covering more than a quarter of the celestial sphere. 
The DR7 is the final data release of the SDSS Legacy Survey. We chose $\sim$ 20,000 galaxy targets ($0.02 < z < 0.06$) from the Main Galaxy Sample and retrieve their images and spectra. 
We then select highly inclined edge-on disk galaxies by utilizing the morphological classification data from the Galaxy Zoo 2 (GZ2) project. 
The morphological classification was done by \citet{2013MNRAS.435.2835W} and we make use of the class ``spiral galaxy other," which consists of highly inclined edge-on spiral galaxies whose spiral arms cannot be distinguished.
When selecting edge-on galaxies we have the following criteria: ($a$) they are flagged as spirals, which requires 80 \% of the volunteer votes for the spiral category after the de-biasing procedure, and ($b$) they are classified as edge-on spirals for which more than a half of the volunteers voted for the spiral category. 
Many of the sample galaxies are too small and/or too faint to identify the presence of the warped structure. 
We only use galaxies large and massive enough to analyze their warped disk structures. 
We select $\sim$11,000 edge-on disk galaxies with $g$-band isophotal major axis $A_{\rm iso} > 22''$ and stellar mass $M_{*}/M_{\odot}$\,$>$\,$10^9$. 
Through visual inspection, we further remove $\sim$3000 galaxies that are not suitable to measure their warped structures due to their intricate dust lanes, spiral arms, and overlapped stars/galaxies at the edge of the galaxies.

The stellar masses are taken from the \citet{2014ApJS..210....3M} catalog that provides photometric data of one million galaxies in the SDSS. 
In the catalog, the stellar masses for bulges, disks, and total are based on the updated bulge + disk decomposition data from \citet{2011ApJS..196...11S}. 
In \citet{2011ApJS..196...11S}, they used a successful bulge + disk fitting code, {\fontfamily{qcr}\selectfont GIM2D}. 
The size measurements from {\fontfamily{qcr}\selectfont GIM2D} are converted into the stellar masses for the decomposed bulge and disk components separately using all $ugriz$ wavebands. 
However, \citet{2014ApJS..210....3M} cautioned that in the case of highly inclined edge-on galaxies, the total stellar mass and the sum of decomposed bulge + disk stellar mass can differ. 
We select only the reliable galaxies that are within the five times of standard deviation from the correlation between the total stellar masses and the sum of bulge + disk masses.

The optical colors and sSFR in this study are taken from the MPA/JHU catalog (\citealt{2004MNRAS.351.1151B}) that provides spectroscopic data of galaxies in the SDSS. The MPA/JHU catalog lists optical colors as well as sSFR estimated from emission lines such as H$\alpha$, H$\beta$, [OII], [NII], and [SII]. To examine the sSFR enhancement across a galaxy, we use the aperture corrected sSFR data. Galaxies that show emission lines of AGNs are also removed. The AGN candidates are selected using the distribution on the BPT diagram  (\citealt{1981PASP...93....5B}; \citealt{2003MNRAS.341...33K}). We exclude `AGN' and `composite' galaxies. We note that this procedure can not classify $\sim$700 galaxies with feeble emission lines because they are not shown on the BPT diagram. They are essentially normal quiescent galaxies with no AGN activity and included in our sample. Our sample contains $\sim$8000 non-AGN, edge-on galaxies.

\subsection{Measurements of the Warped Stellar Disks}

We retrieve the corrected frames of our sample galaxies from the SDSS Data Archive Server. The corrected frame is an imaging frame from the SDSS imaging pipeline that has been bias-subtracted, flat-flagged, and pixel-defect corrected. We crop the field images based on the galaxy positions on the frame. The resultant images used in our analysis are $500 \times 500$ pixel ($3.3' \times 3.3'$) FIT images centered on the sample galaxies. In this study, we use the $g$-band images. The signal-to-noise, sensitivity for foreground stars, and dust extinction depend on wavebands, but we do not find any dependence of measured warp properties on wavebands such as $u$-, $r$-, $i$-, and $z$-band.

In order to extract the overall shape of disks, we blur the images of the target galaxies using the {\fontfamily{qcr}\selectfont SMOOTH} function. This procedure returns a copy of array smoothed with 5-pixel width. Thus, after the smoothing procedure, the scale of our galaxy images becomes five times smaller ($100 \times 100$ pixel). The center of our target galaxy is defined as the location of the brightest point in the smoothed image. The sky is not subtracted in the SDSS imaging pipeline and we thus subtract the background sky values from the original images. The background sky is measured as the median pixel value outside of the region of interest.

\begin{figure}
	\includegraphics[width=\columnwidth]{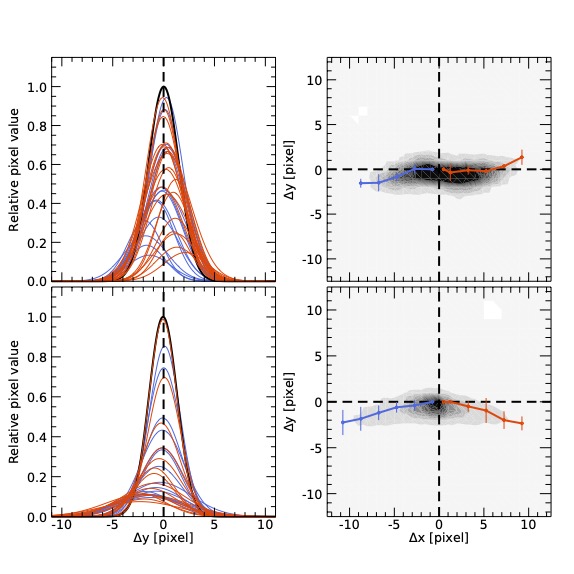}
    \caption{The procedure of the warp angle measurement through our automated scheme for the case of S-type (upper row) and U-type warps (lower row). 
    (Left column) The Gaussian fitted vertical distribution of pixel values from the left side of the disk (blue) via the center peak (black) to the right side (red) with respect to the major axis. 
    (Right column) The warped structure of the `spine' is illustrated by blue (left side) and red (right side) solid lines, respectively, on the contour image of each galaxy. 
    The length of the error bars is the standard error of the location of the spine at each $\Delta$x bin.}
    \label{fig:1}
\end{figure}

To quantify the warp structure, we newly invent an automated warp measurement scheme. 
Figure~\ref{fig:1} illustrates the procedure of the warp angle measurement through our automated scheme for the case of S-type (upper row) and U-type warps (lower row). 
Specifically, we first align the major axes of galaxies horizontally based on the position angle (PA) from SDSS DR7 database, which is defined as the angle with respect to the north celestial pole following the direction of the right ascension. 
We then derive the vertical brightness distribution using a Gaussian function of the vertical distance from the major central axis at each x-coordinate from the left to the right side. 
In this procedure, we only use pixels whose value is brighter than five times the standard deviation of the background sky value. 
We join all peak points of the vertical brightness distribution at each x-coordinate as a single curved line, which are defined as the `spine' of target galaxy.
To derive the central major axis, we apply the orthogonal linear regression to the central spine points within the half of disk size.

The estimated major axis is usually misaligned with the horizontal line due to an inaccurate PA from the SDSS pipeline data, and even after the first rotation of our target images, the PA is not zero. 
This misalignment impedes the measurement of the exact value of warping amplitudes. Thus, we determine the major axis and warped disk structure again by repeating the rearrangement procedure until the calculated PA meets the tolerance (PA $< 0.01^{\circ}$). 
Finally, the warping amplitude, $\alpha$, is calculated as the degree of misalignment between the central major axis and tips at the outermost bent structures. 
The warping amplitudes are measured on both sides of the galactic disk. 
To avoid spurious warp detection, we only consider a disk whose vertical deviation at the outermost point is more significant ($>3\sigma$) than the fluctuation of wobbling along the central major axis.
Between the warping amplitudes of the two sides, we take the larger one as the major warping amplitude, $\alpha$. 
We then divide the morphology of disks into S-type, U-type, and unwarped based on the degree and direction in which each endpoint of the disk is bent (\citealt{1998A&A...337....9R}; \citealt{2002A&A...382..513R}; \citealt{2003A&A...399..457S}; \citealt{2006NewA...11..293A}).

\subsection{Warped Disk Galaxy Sample}

\begin{figure}
	\includegraphics[width=\columnwidth]{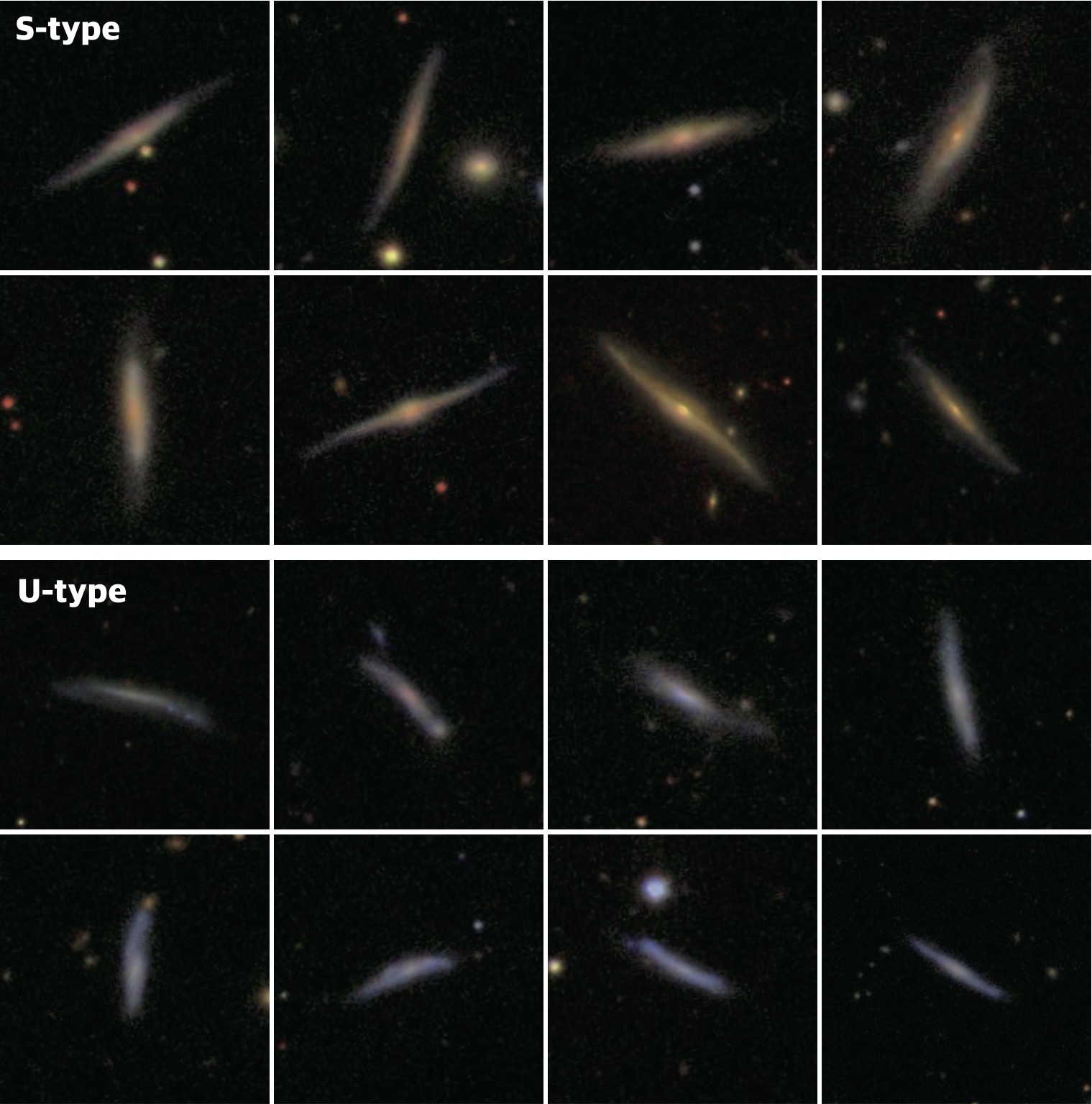}
    \caption{Randomly selected sample of multi-band SDSS images of S- (upper eight panels) and U-type warped (lower eight panels) galaxies among our final catalog.}
    \label{fig:2}
\end{figure}

In this work, we present a new statistical analysis of optical warps of disk galaxies with a much larger body of data than has been used in previous studies (\citealt{1998A&A...337....9R}; \citealt{2002A&A...382..513R}; \citealt{2003A&A...399..457S}; \citealt{2006NewA...11..293A}; \citealt{2016MNRAS.461.4233R}).
Using our automated warp measurement scheme, after all pre-procedures, we identify 3662 warped disk galaxies out of $\sim$8000 highly inclined edge-on galaxies in the local universe ($0.02 < z < 0.06$). 
Among them, we have 2206 S-types and 1456 U-types. 
Table~\ref{tab:example_1} gives the basic properties of our S- and U-type warped galaxies, including the number of galaxies, number fraction, median warp angle, and median stellar mass. 
The properties are in agreement with the aforementioned studies. 
In particular, warps are very common with the fraction of $\sim$50 \% of edge-on disk galaxies, and S-types are about 1.5 times more frequent and exhibit slightly stronger warping amplitudes than U-types.
Figure~\ref{fig:2} shows some randomly selected example images of S- and U-type warps among our final catalog.

\subsection{Control Sample}

\begin{figure}
	\includegraphics[width=\columnwidth]{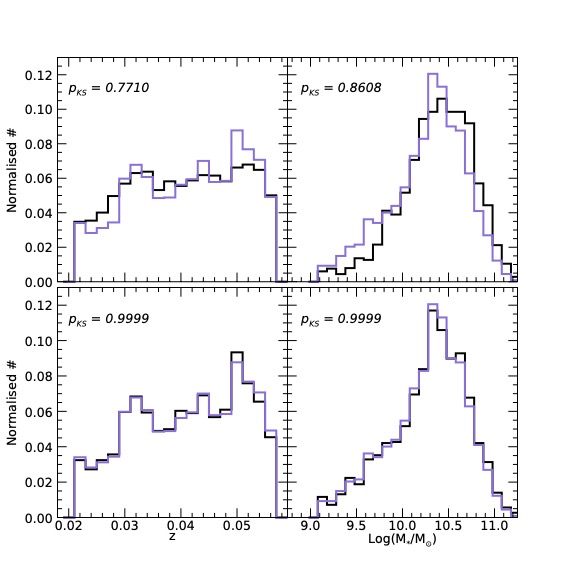}
    \caption{(Upper) Distributions of the redshift (left column) and stellar mass (right) for unwarped galaxies (black histogram) and warped galaxies (violet). 
    For each normalised histogram, the integral under the histogram is equal to one. 
    Each p-value by the KS test is given on each panel. 
    (Lower) The same as the upper panels, but black histograms are for the control sample.}
    \label{fig:3}
\end{figure}

The upper panels of Figure~\ref{fig:3} show the distributions of redshift and stellar mass of unwarped and warped galaxies. Compared to the unwarped galaxies, the warped galaxies exhibit slightly higher redshift and lower stellar mass.
The larger redshift for warped galaxies is likely due to the selection bias of warp detection. 
Because of their closer distance and thus larger apparent size, there is higher chance that target galaxies are overlapped with other stars and/or galaxies. 
The different distribution of stellar mass between warped and unwarped galaxies can be explained by the mass dependence on external mechanisms. 
Simulations of galaxy--galaxy interactions showed that the extent of the morphological changes depends on the mass ratio of interacting galaxies, in the sense that less massive galaxies are more susceptible to external forces (e.g., \citealt{2008MNRAS.384..386C}; \citealt{2017MNRAS.465.3446G}).

To avoid the effect of different redshift and stellar mass between warped and unwpared galaxies, we carefully construct a control sample of unwarped edge-on galaxies.
We randomly select an unwarped galaxy for each warped galaxy within the bin range of $\pm$ 0.005 in redshift and $\pm$ 0.1 dex in stellar mass.
In the lower panels of Figure~\ref{fig:3}, the control sample exhibits redshift and stellar mass distributions nearly identical to those of warped galaxies.

\section{Physical Properties of Warped Disk Galaxies}
\label{sec:result}

To address which mechanism governs the formation of the different warp morphologies, we begin with comparison of some key properties of S- and U-type warped galaxies with respect to those of the unwarped control sample, including optical colors, sSFR, and environment.

\subsection{The Optical Color and Star Formation Rate}

\begin{figure}
	\includegraphics[width=\columnwidth]{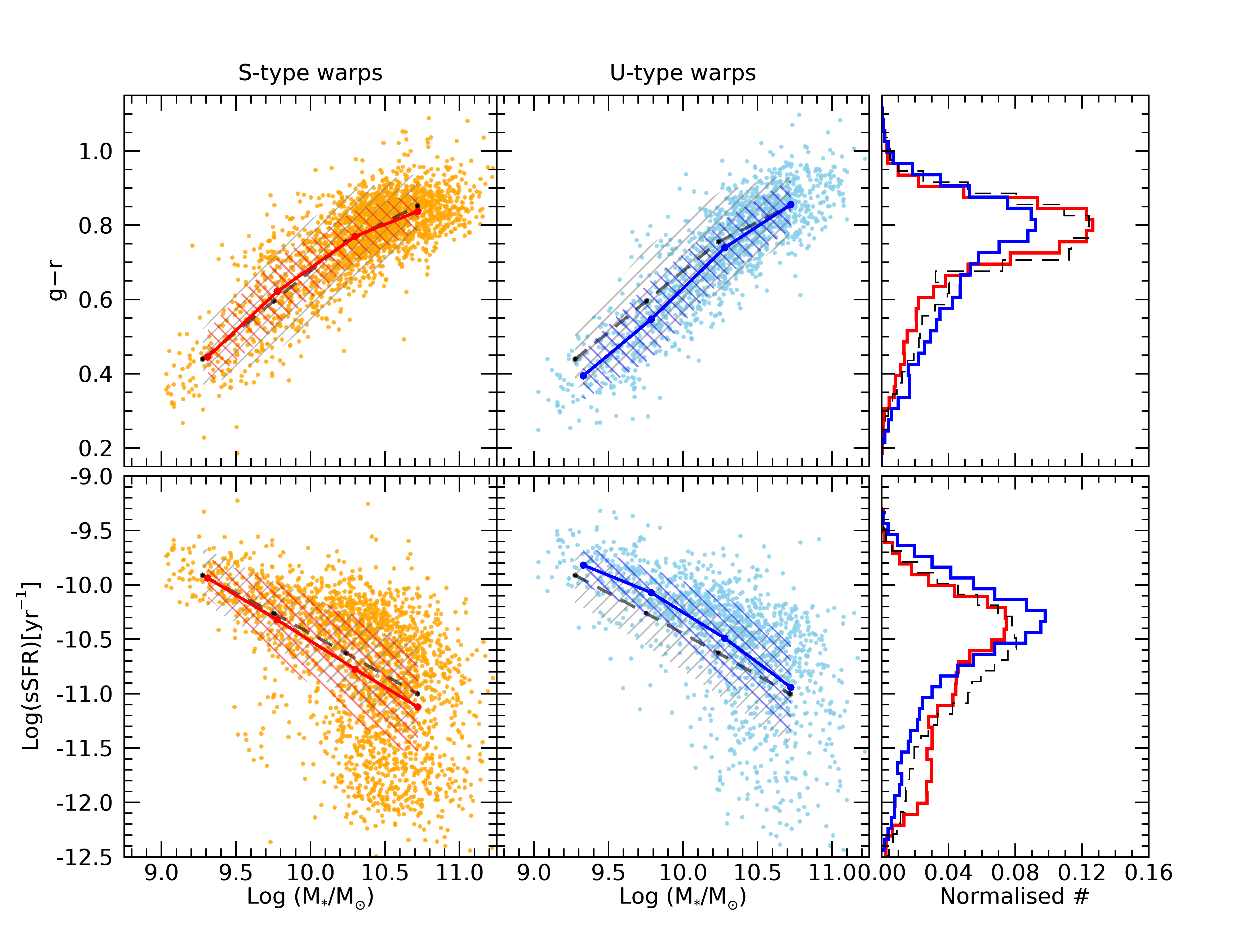}
    \caption{(Upper, left two panels) The $g-r$ colors versus stellar mass for S-types (orange dots on the left panel) and U-types (cyan dots on the right). The $g-r$ color is reddening- and k-corrected.
    Red, blue solid and black dashed lines connect the median values of the Log($M_{*}/M_{\odot}$) bins for S-type, U-type warps and unwarped control galaxies, respectively. 
    Black, red and blue dashed regions are standard deviation scatter.
    The length of the error bars is the standard error, which is estimated as the standard deviation divided by the square root of the sample size at each mass bin. 
    (Upper, rightmost panel) Each normalised histogram shows the color distribution of S-types (red solid histogram) and U-types (blue solid), the unwarped control sample (black dashed). 
    (Lower) The same as the upper panels, but for the Log(sSFR).}
    \label{fig:4}
\end{figure}

In Figure~\ref{fig:4}, the upper panels show the distribution of SDSS optical $g-r$ colors of warped and unwarped galaxies as functions of stellar mass. 
Interestingly, we discover an unexpected discrepancy between S- and U-type warps in optical $g-r$ colors. 
While S-type warped galaxies show $g-r$ similar to unwarped control sample, U-types are buler by $\sim$0.05 dex.
In the lower panels, we also find that U-type warps exhibit $\sim$0.25 dex higher sSFR than unwarped control galaxies. 
The blue-ward color offset and the increase of sSFR are greater for less massive galaxies ($M_{*}/M_{\odot}<10^{10}$).

\begin{figure}
	\includegraphics[width=\columnwidth]{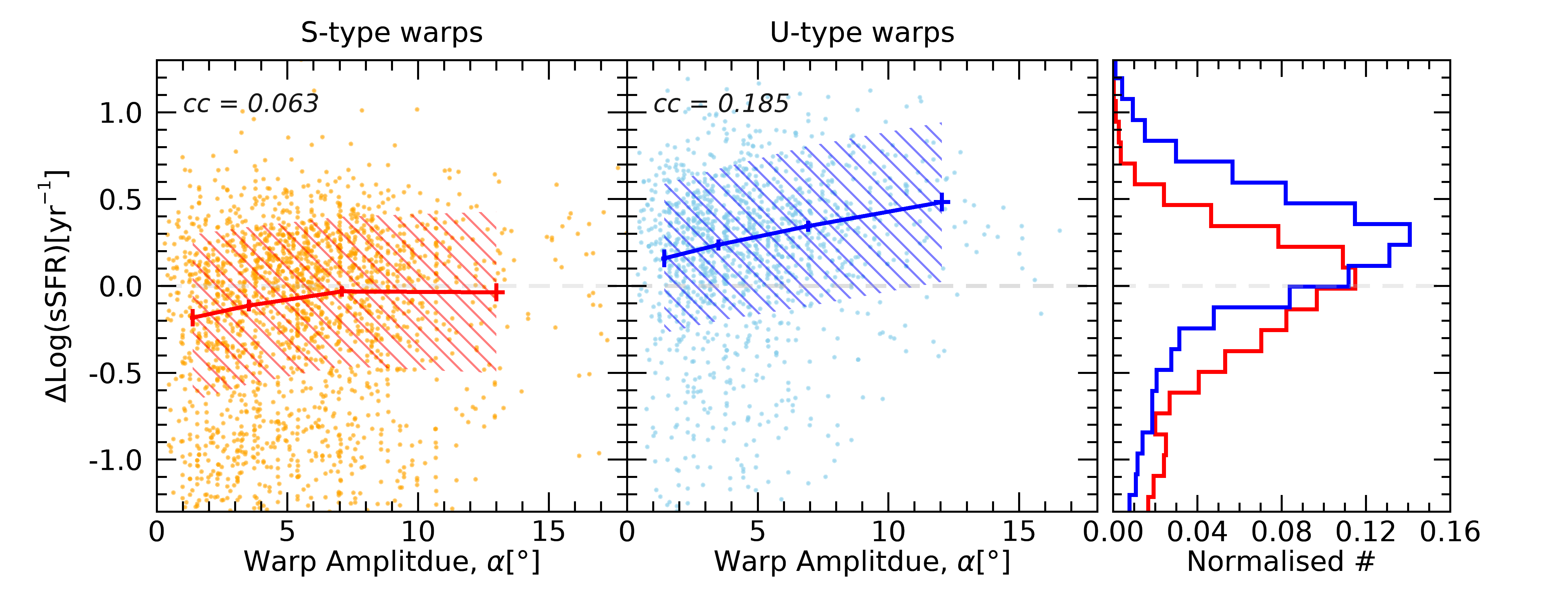}
    \caption{(Left two panels) The difference in sSFR between the warp sample and the mean sSFR of its corresponding control sample, $\Delta$Log(sSFR), as a function of warping amplitude, $\alpha$, for S-types (orange dots on the left panel) and U-type (cyan dots on the left panel). 
    Red and blue lines connect the median $\Delta$Log(sSFR) values of the warp amplitude bins for S-types and U-types, respectively.
    Red and blue dashed regions are standard deviation scatter.
    The length of the error bars is the standard error, which is estimated as the standard deviation divided by the square root of the sample size at each $\alpha$ bin.  
    For comparison, $\Delta$Log(sSFR) = 0.0 is marked by horizontal grey dashed line.
    (Rightmost panel) Each normalised histogram shows the $\Delta$Log(sSFR) distribution of S-types (red solid histogram) and U-types (blue solid). }
    \label{fig:5}
\end{figure}

In Figure~\ref{fig:5}, we compare the residuals of sSFR as a function of the warping amplitude. 
We define the residual of sSFR, $\Delta$Log(sSFR), as the difference in sSFR between the warp sample and the mean sSFR of its corresponding stellar mass- and redshift-matched control sample in bins of stellar mass.
Pearson correlation coefficient (cc) is shown in each panel of Figure~\ref{fig:5}.
There is a distinct correlation between $\Delta$Log(sSFR) and warping amplitudes only for U-type warps with cc = 0.185.
Strongly warped ($\alpha > 10^{\circ}$) U-type galaxies have $\sim0.3$ dex higher sSFR than weakly warped ($\alpha < 5^{\circ}$) U-type ones. 
The enhancement of sSFR for strongly warped U-type galaxies supports that the galaxies are associated with more efficient star formation activity than S-types and unwarped galaxies. 
This discrepancy between S- and U-type warps implies that the two morphologies are probably governed by distinct mechanisms.

\subsection{The Environment}
\label{sec:Env}

Conventional simulations of galaxy-galaxy interactions suggested that S-type warps are common products of tidal interactions (\citealt{2014ApJ...789...90K}; \citealt{2017MNRAS.465.3446G}; \citealt{2018MNRAS.481..286L}; \citealt{2018Natur.561..360A}). 
However, whether formation of U-type warps is governed by the same mechanism is unclear. 
To explore the environmental effects on galactic warps, we examine the incidence and amplitudes of S- and U-type warps as functions of environmental parameters from the local to galaxy groups/clusters scale.

\subsubsection{The Local Environmental Effect}

We investigate the effect of the local environment on warp formation by examining whether the frequency and amplitude of warps depend on environmental parameters.
We define two different local environmental parameters: ($a$) local density of the surrounding area, $\Sigma_{\textrm{N}}$, ($b$) tidal influence of the nearest neighboring galaxy, $F_{\textrm{tidal}}$. 
To estimate the local density, we adopt the projected surface density used by \citet{2006MNRAS.373..469B}, such that \begin{equation} \Sigma_{\textrm{N}} = \cfrac{N}{\pi {d_\textrm{N}}^2} \hspace{1mm}, \end{equation}where $d_{\textrm{N}}$ is the comoving distance to the N$^{th}$ nearest galaxy. The projected surface density Log($\Sigma_{\textrm{45}}$) is used for our study, which is defined as the average value of Log($\Sigma_{\textrm{N}}$) for N = 4 and 5 as proposed by \citet{2006MNRAS.373..469B}. 
We also use $F_{\rm tidal}$ defined as \begin{equation} F_{\textrm{tidal}} = \text{Log}(M_{*}/M_{\odot}) - 3\,\text{Log}(r_{\textrm{nearest}}) \hspace{1mm}, \end{equation}where $M_{*}$ is the stellar mass of the closest neighboring galaxy and $r_{\textrm{nearest}}$ is the distance to the neighboring galaxy in kpc. 
We define the nearest neighbor as a galaxy that is of the closest projected distance within a radial velocity difference of 1000 km\, s$^{-1}$ and with mass between 0.1 and 10 of the target galaxy.

\begin{figure}
	\includegraphics[width=\columnwidth]{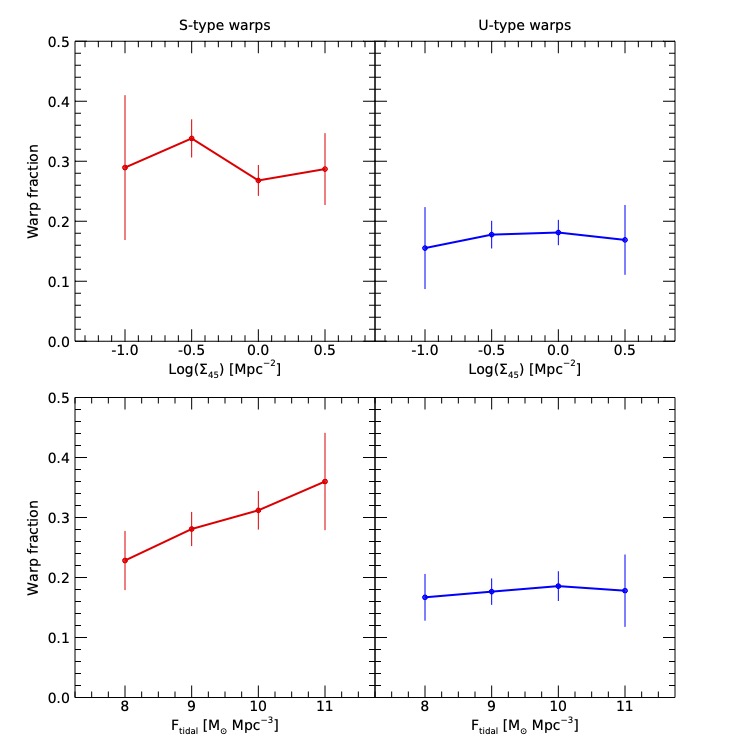}
    \caption{(Upper panels) Warped galaxy fraction of S-types (red line on the left) and U-types (blue line on the right) as a function of local density, Log($\Sigma_{45}$). 
    The length of the error bars is the poission error in each Log($\Sigma_{45}$) bin. 
    (Lower panels) The same as the upper panels, but warped galaxy fraction as a function of tidal force, $F_{\rm tidal}$, by the closest neighbor galaxies.}
    \label{fig:6}
\end{figure}

\begin{figure}
	\includegraphics[width=\columnwidth]{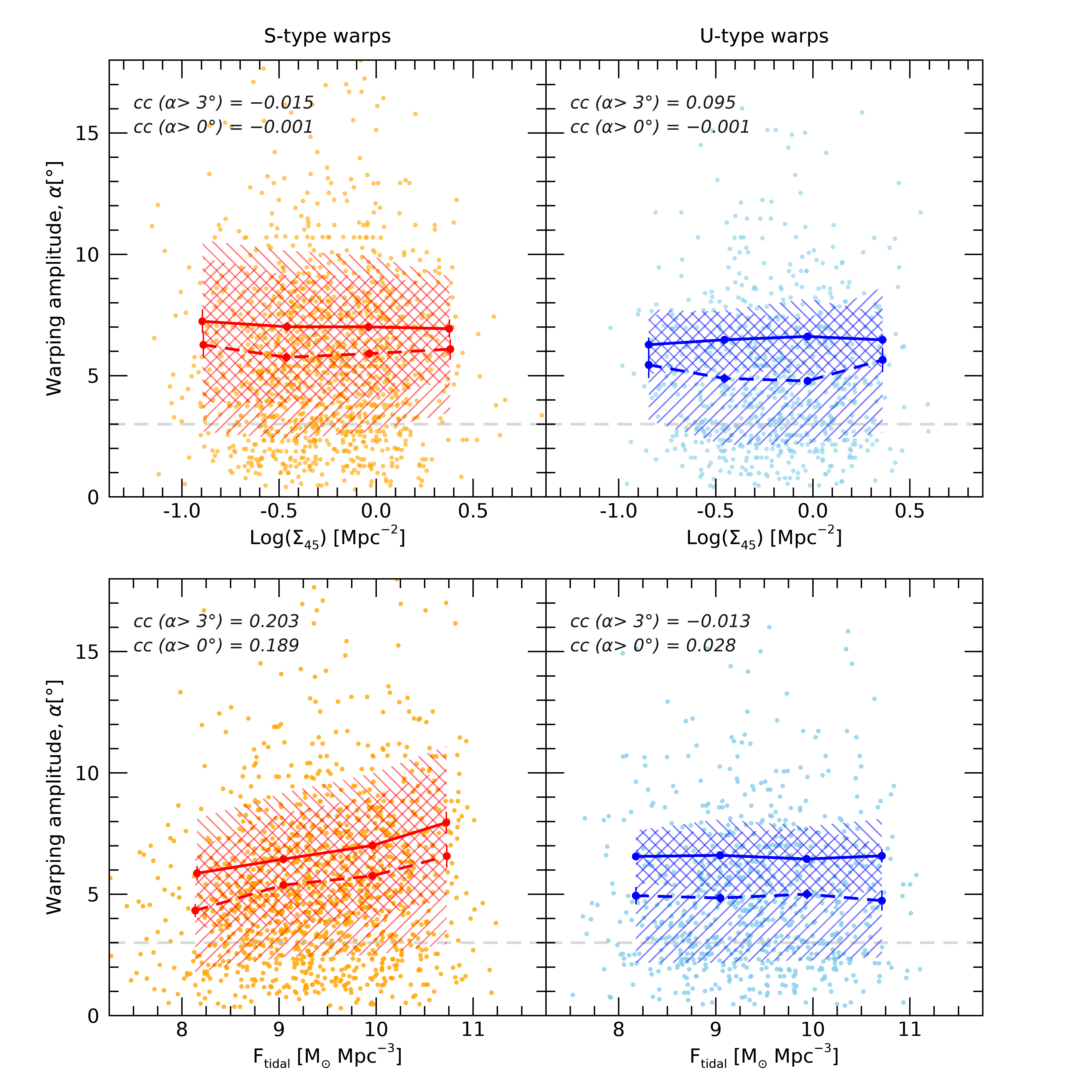}
    \caption{(Upper panels) Distribution on the Log($\Sigma_{45}$) vs. warp amplitude plane for S-types (orange dots on the left panel) and U-types (cyan dots on the right). 
    Red and blue lines connect the median $\alpha$ values of the Log($\Sigma_{45}$) bins for S-types and U-types, respectively.
    Dashed and solid lines follow the median distribution of all and strongly warped ($\alpha > 3^{\circ}$) galaxies, respectively.
    Red and blue dashed regions are standard deviation scatter.
    The length of the error bars is the standard error, which is estimated as the standard deviation divided by the square root of the sample size at each Log($\Sigma_{45}$) bin. 
    We estimated the correlation coefficient (cc) for all warpred and strongly warped galaxies.
    (Lower panels) The same as the upper panels, but for the $F_{\rm tidal}$ vs. warp amplitude plane.}
    \label{fig:7}
\end{figure}

Figures~\ref{fig:6} and \ref{fig:7} show respectively the frequency and amplitude of warped structures as functions of both local environmental parameters, Log($\Sigma_{\textrm{45}}$) and $F_{\textrm{tidal}}$. 
In their upper panels, there is no detectable environmental effect on the warp fraction and amplitude. 
Intriguingly, the lower panels of Figures~\ref{fig:6} and \ref{fig:7} show that S-type warps depend on the tidal force by the nearest neighbor galaxies, $F_{\textrm{tidal}}$.
As the tidal influence increases, the incidence and the warping amplitude of S-warps increases with cc = 0.189. When we consider the strongly warped galaxies ($\alpha > 3^{\circ}$), the correlation becomes slightly more significant with cc = 0.203.
Contrary to S-type warps, the frequency and amplitudes of U-types depend on neither environmental parameters. 
This implies that the tidal interaction plays a role in formation of S- and U-type warps differently. 
This result is consistent with previous theoretical studies, which usually reproduced S-types with few or no U-types by tidal interactions (\citealt{1995ApJ...455L..31W}; \citealt{2000ApJ...534..598V}; \citealt{2006ApJ...641L..33W}; \citealt{2008MNRAS.388..697M}; \citealt{2013MNRAS.429..159G}; \citealt{2014ApJ...789...90K}; \citealt{2017MNRAS.465.3446G}; \citealt{2018MNRAS.481..286L}; \citealt{2013MNRAS.429..159G}; \citealt{2018Natur.561..360A}).

\subsubsection{The Group/Cluster Effect}

We examine the effect of the group/cluster environment on galactic warps. 
We select the member galaxies of groups and clusters using the SDSS DR8 group/cluster member galaxy catalog by \citet{2012A&A...540A.106T}. 
The group/cluster effects such as RPS and harassment are considered to happen usually in massive clusters ($M_{\textrm{halo}}/M_{\odot}$\,$>$\,$10^{14}$). 
However, some recent observations reported that RPS galaxies exist even in low-mass galaxy groups (e.g., \citealt{2011atnf.prop.3970W}; \citealt{2012ApJ...757..122R}; \citealt{2018MNRAS.480.3152V}; \citealt{2019MNRAS.487.2797E}).
\citet{2011MNRAS.416.3170T} determined the RPS effect in groups and clusters with the halo mass range of $10^{12.5}$\,$<$\,$M_{\textrm{halo}}/M_{\odot}$\,$<$\,$10^{15.35}$. They concluded that the strength of RPS becomes more prominent as the halo mass increases, but galaxies in low-mass groups also experience mass loss by RPS. 
\citet{2016AJ....151...78P} reported 344 jellyfish candidates in 71 galaxy clusters and 75 candidates in lower mass groups ($10^{11}$\,$<$\,$M_{\textrm{halo}}/M_{\odot}$\,$<$\,$10^{14}$). 
Although the exact physical origin of jellyfishes in group environments should be securely investigated, the authors introduced convincing cases of jellyfish galaxies in group and lower mass halos. 
They stated that \textit{``jellyfish galaxies could be present even in groups and lower mass halos."}
\citet{2021A&A...652A.153R} also identified 60 jellyfish galaxies in $\sim$500 galaxy groups with halo mass of $10^{12.5}$\,$<$\,$M_{\textrm{halo}}/M_{\odot}$\,$<$\,$10^{14}$ and compared them with LOFAR Two-metre Sky Survey (LoTSS) jellyfish galaxies.

\begin{figure}
	\includegraphics[width=\columnwidth]{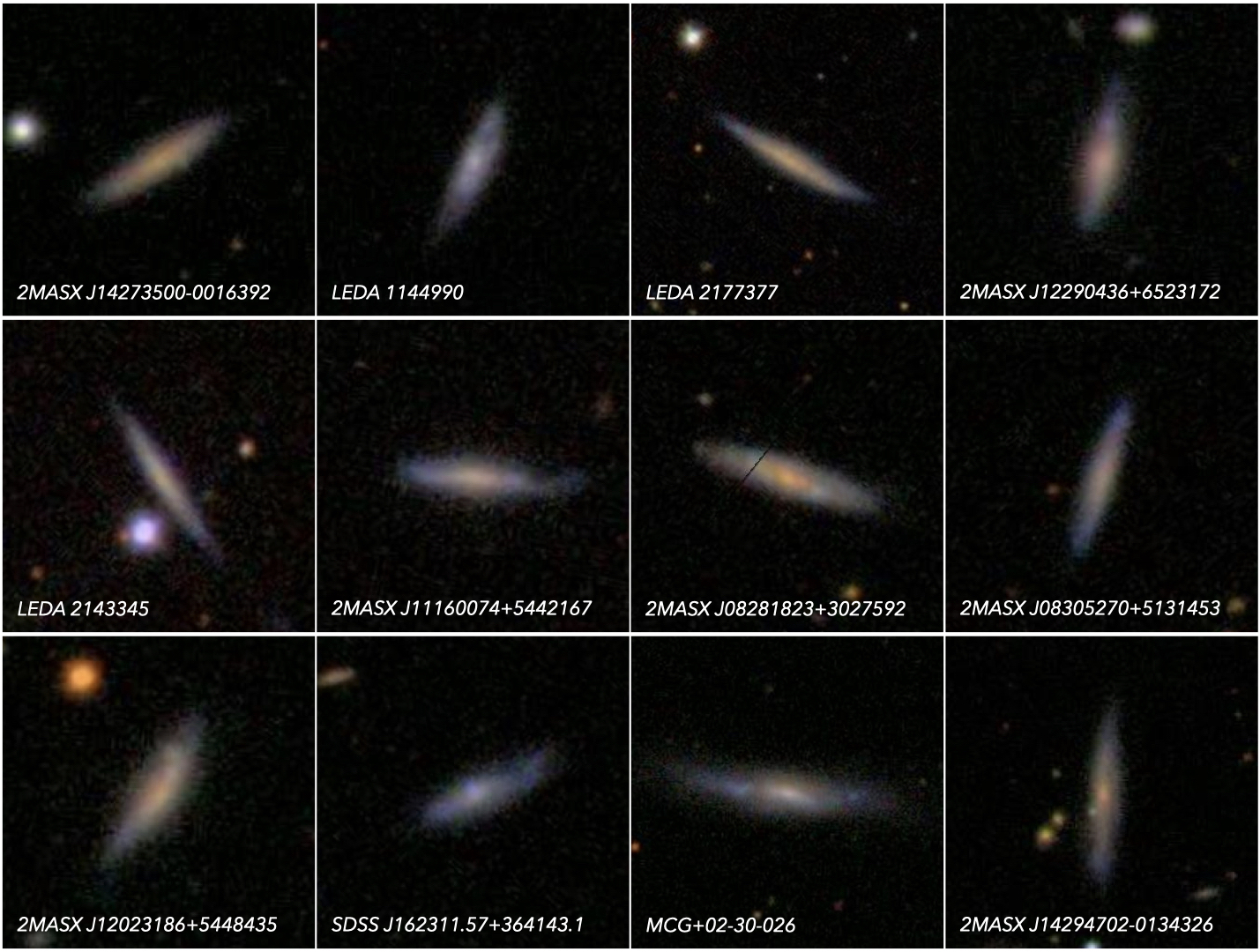}
    \caption{Best sample of multi-band SDSS images of U-type warped galaxies in massive clusters with their name.}
    \label{fig:8}
\end{figure}

Adopting the recent results, we use two different definition of group/cluster environments: ($a$) a loose criterion including galaxy groups of low halo mass ($M_{\textrm{halo}}/M_{\odot}$\,$\geq$\,$10^{12}$), ($b$) a conventional tight criterion only including massive galaxy clusters ($M_{\textrm{halo}}/M_{\odot}$\,$\geq$\,$10^{14}$). 
When we adopt the looser criterion, 13.2\% of S-types and 17.5\% of U-types belong to groups.
When using the conventional criterion, 2.2\% of S-types and 4.1\% of U-types are classified as cluster member galaxies. 
Figure~\ref{fig:8} shows best example images of U-type warped galaxies in massive clusters. While only a small fraction of warped galaxies are involved in groups/clusters, U-type warps are slightly more common in these environments. 
However, since the tidal and group/cluster environmental effects occur simultaneously, we find no explicit dependency of warp fractions and warping amplitudes on host halo masses.

\begin{figure}
	\includegraphics[width=0.9\columnwidth]{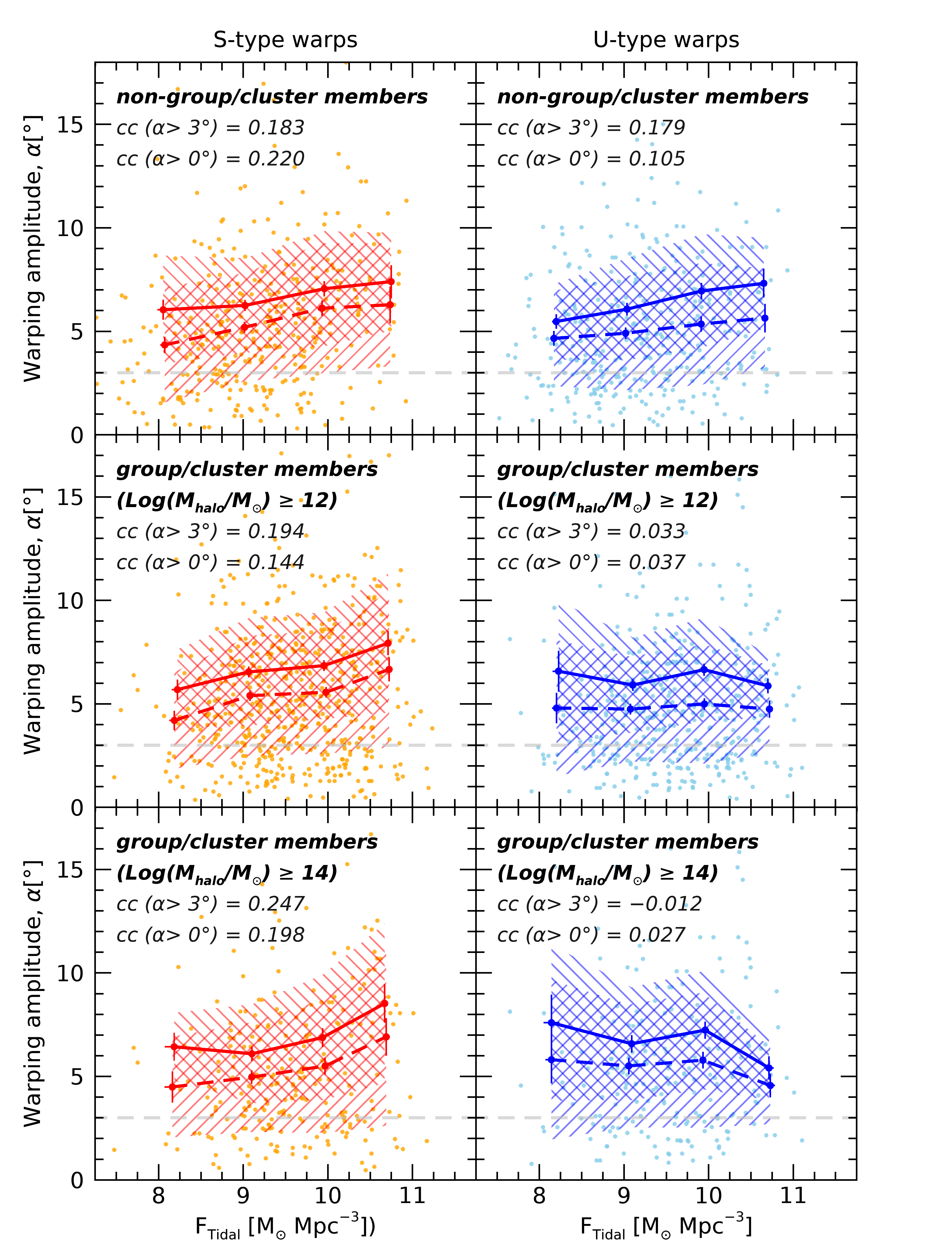}
    \caption{The same as the lower panels of Figure~\ref{fig:8}, but for non-group/cluster members (upper panels), group/cluster members with host halo masses of $M_{\textrm{halo}}/M_{\odot}$\,$\geq$\,$10^{12}$ (middle), and group/cluster members with host halo masses of $M_{\textrm{halo}}/M_{\odot}$\,$\geq$\,$10^{14}$ (lower).}
    \label{fig:9}
\end{figure}

To distinguish the effect of tidal interactions and group/cluster environments, we examine the tidal effects for non-group/cluster and group/cluster galaxies separately.
In Figure~\ref{fig:9}, the upper panels show the warping amplitudes of S- and U-type warps in non-group/cluster environments as a function of $F_{\textrm{tidal}}$.
The tidal force is exerted by the nearest neighbor and thus isolated warped galaxies are not shown in this figure.
The middle and lower panels show warped galaxies in groups and clusters.
We exclude the central galaxy in each group/cluster to examine the the group/cluster effect on infalling galaxies only.
S-type warps in non-group/cluster and group/cluster environments show similar positive correlations with cc = 0.220.
This implies that tidal interactions is important in formation of S-types even in groups/clusters.
By contrast, U-types in groups/clusters show no correlation between warping amplitudes and $F_{\textrm{tidal}}$ with relatively low cc = 0.037 and 0.027.
This can be explained by the fact that the imprints of tidal interactions on formation of U-types in groups/clusters are vanished by non-tidal effects such as RPS. 
Thus, it is necessary to investigate how non-tidal mechanisms are at work in construction of U-types in groups/clusters.

\section{Are U-type warps in groups/clusters jellyfishes?}
\label{sec:kine}
   
RPS in groups/clusters often produces galaxies with disturbed HI gas disks and tentacle-like structures, which are so-called ``jellyfish galaxies" (e.g., \citealt{2010MNRAS.408.1417S}; \citealt{2014ApJ...781L..40E}; \citealt{2014MNRAS.445.4335F}; \citealt{2016AJ....151...78P}; \citealt{2017ApJ...844...48P}; \citealt{2018MNRAS.476.3781R}; \citealt{2018MNRAS.476.4753J}; \citealt{2019MNRAS.483.1042Y}).
Some simulations suggested that jellyfish galaxies can exhibit U-shaped \textit{stellar} disks during RPS, specifically at the very early stage. For example, \citet{2012MNRAS.420.1990S} simulated that drag force on the gas disk during RPS can be transmitted to the dark matter halo, and stellar disk can be changed into a U-shape appearance briefly ($<$ 200 Myr).
It is noteworthy that, according to this simulation, U-shaped stellar disks in jellyfishes are bent into the opposite direction of their stripped gas trails. 
This theoretical expectation is consistent with the recent simulation of \citet{2022arXiv220101316L}, in which a U-shaped stellar warp opposite to stripped gas tails is present at the beginning of RPS ($t_\textrm{form} \sim$ 185 Myr).

Motivated by the previous work and our results on U-type warps in groups/clusters not showing the tidal effect in local environments, we investigate whether U-type warps in groups/clusters are related to jellyfish galaxies.
In this section, we discuss the possible link between U-type warps in groups/clusters and jellyfish galaxies, considering their similarities in optical morphology, kinematics in groups/clusters, sSFR, and HI fraction.

\subsection{Warped Jellyfish Galaxies} 

We first look into the appearance of the warped stellar component of RPS jellyfish galaxies in the literature.
Although previous observational studies on jellyfish galaxies did not delve into galaxies' warped disk structures, we find some interesting examples that show detectable U-shaped stellar disks through our visual inspection on their optical images.
Our examples include MACSJ0451-JFG1, MACSJ0712-JFG1, MACSJ1752-JFG1 in \textit{HST} F606W and F814W from \citet{2014ApJ...781L..40E},  A1758N-JFG1 in \textit{HST} F606W and F814W from \citet{2019ApJ...887..158K}, and JO113 in $g$-, $r$-, and $i$-bands from \citet{2020ApJ...899...13G}.
We find other examples from \citet{2020MNRAS.495..554R}; at least five U-shaped stellar warps among eight edge-on galaxies with no S-shaped ones.
\citet{2021NatAs...5.1308G} investigated the effect of RPS on low-mass galaxies in the Coma and Abell 2147 clusters. We find through our visual inspection that at least four out of five edge-on galaxies (GMP 3176, GMP 2639, GMP 4348, and J160231.45+155749.9) are bent into U-shape morphologies.
However, since these observations did not include radio wavelengths, it is limited to directly compare the directions of stellar warps and stripped gas tails.

Thanks to other recent multi-wavelength observations, we find further promising examples consistent with the theoretical expectation by \citet{2012MNRAS.420.1990S}.
\citet{2021arXiv211104501M} identified 13 jellyfish galaxies from Abell 2744 and Abell 370 clusters. 
They provided RGB images from HST of jellyfish galaxies with overlapped [OII] emission from MUSE observation. 
We find two edge-on galaxies, A370-06 and A370-08. 
Galaxy A370-06 shows a U-shaped stellar disk bent in the opposite direction of gas tails.
\citet{2021A&A...652A.153R} provided $\sim$60 jellyfish galaxies' $g$-band optical images with overlapped LOFAR 144MHz maps.  
Through our visual inspection, we find one S-shaped stellar disk (KUG0930+342) and two U-shaped disks (LEDA2158637 and LEDA2157975) among 20 edge-on galaxies from their sample. 
Specifically, LEDA2157975 exhibits a U-shaped stellar disk clearly bent in the opposite direction of stripped gas components.

\begin{figure}
	\includegraphics[width=0.9\columnwidth]{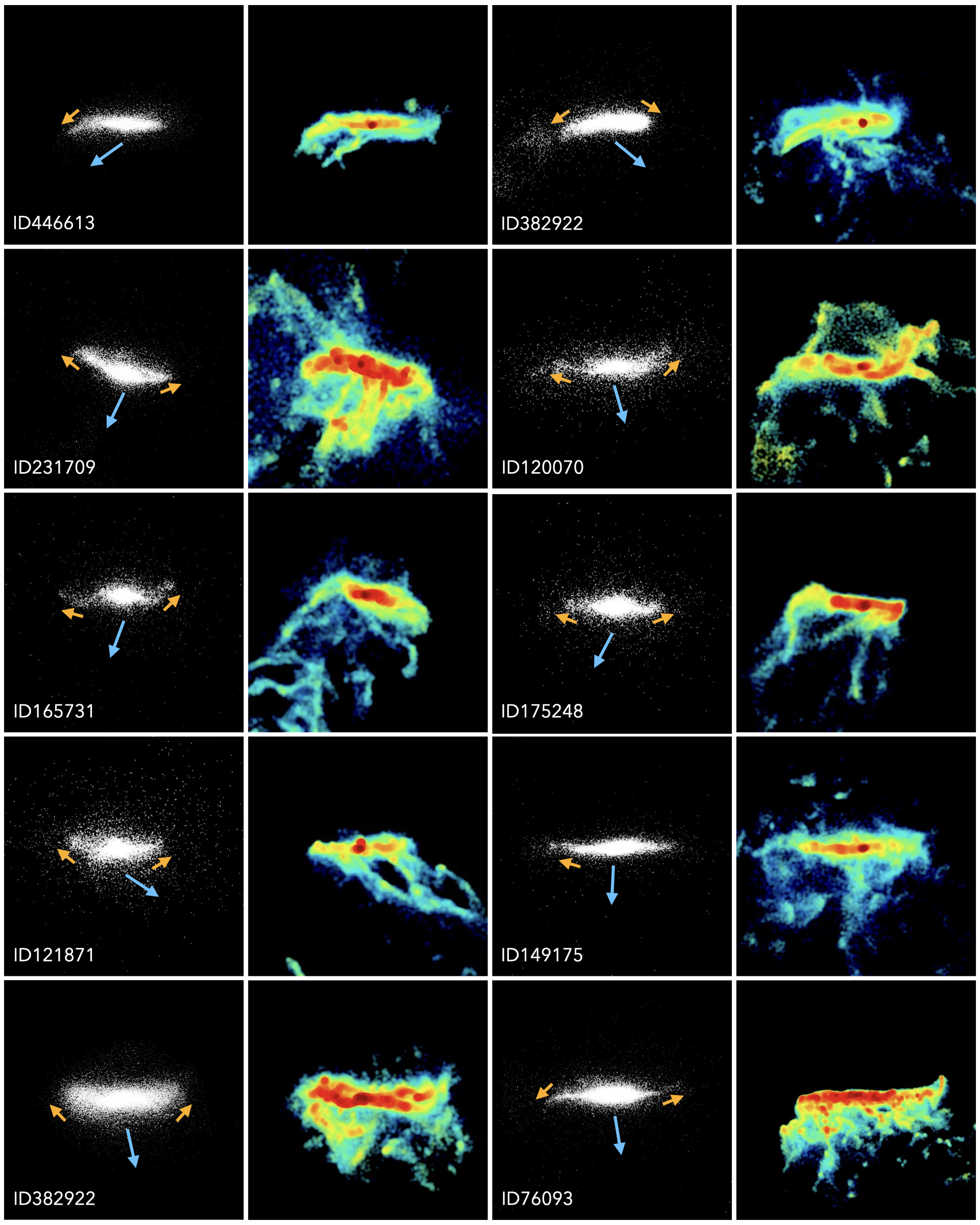}
    \caption{Example images of warped stellar disks (1st and 3rd columns) and their gas components (2nd and 4th columns) for RPS-induced jellyfish galaxies from IllustrisTNG simulation. 
    The galaxies are classified as jellyfishes by visual inspection by \citet{2019MNRAS.483.1042Y} and their ID's are given at the lower left corner of the 1st- and 3rd-column panels. 
    The directions of warped stellar disks and those of corresponding gas tails are denoted as orange and cyan arrows, respectively. }
    \label{fig:10}
\end{figure}

We further investigate morphologies of stellar disks of jellyfish galaxies in the Illustris TNG100 simulation (\citealt{2018MNRAS.475..648P}; \citealt{2019ComAC...6....2N}). \citet{2019MNRAS.483.1042Y} identified 800 candidates of jellyfish galaxies with host halo mass of $10^{13}$\,$<$\,$M_{\textrm{halo}}/M_{\odot}$\,$<$\,$10^{14.6}$ through visual inspection.
Figure~\ref{fig:10} illustrates best example images of stellar and gas components of the jellyfishes.
We find 50 jellyfish galaxies that show detectable U-shaped stellar disks and only one example of the S-shape disk.
The direction of warped stellar disks and gas tails are marked as orange and cyan arrows, respectively.
Most U-shaped jellyfish galaxies show detectable U-type stellar warps bent in the opposite direction of their stripped gas tails.
This is consistent with the numerical expectation of \citet{2012MNRAS.420.1990S} and \citet{2022arXiv220101316L}.

We note that there is one case of the S-shaped stellar warp (ID76093) among Illustris TNG100 jellyfish candidates.
The galaxy has the S-shaped gas component; thus it seems that not only RPS but other gravitational mechanisms influence both its stellar and gaseous disks. 
Still, it is necessary to trace how the morphologies of warped disks evolve and prove whether the RPS govern their appearances. 
We will investigate the time-dependent evolution of S- and U-type warps using simulations in forthcoming papers.

\subsection{The Phase-Space Distribution of Group/Cluster Galaxies}

\begin{figure}
\includegraphics[width=0.7\columnwidth]{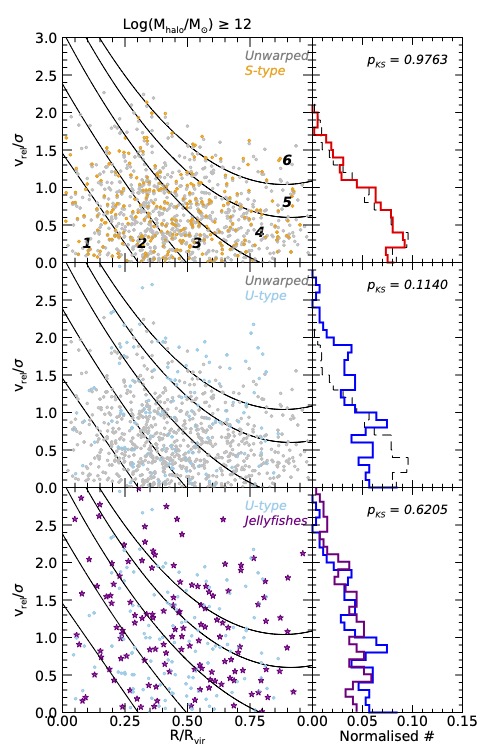}
    \caption{(Top left panel) Distribution of S-type warped (orange dots) and unwarped (grey dots) galaxies in groups/clusters on the $R/R_{\rm vir}$--$v_{\rm rel}/\sigma$ phase-space diagram. 
    Black loci are the demarcation lines given by \citet{2019MNRAS.484.1702P} and the regions ($R/R_{vir} < 1.0$) are labelled as Region 1 through 6.  
    (Top right panel) The normalised $v_{\rm rel}/\sigma$ distributions are illustrated for S-types (red solid histogram) and unwarped (black dashed), along with the p-value from the KS test. 
    (Middle) The same as the top panel, but for U-type warped galaxies (cyan dots on the left and blue solid histogram on the right). 
    (Bottom) The same as the top panel, but for U-type warped galaxies (cyan dots on the left and blue solid histogram on the right) and jellyfish candidates (\citet{2016AJ....151...78P} and \citet{2018MNRAS.476.4753J}; purple stars on the left and purple histogram on the right).}
    \label{fig:11}
\end{figure}

\begin{figure}
\includegraphics[width=0.7\columnwidth]{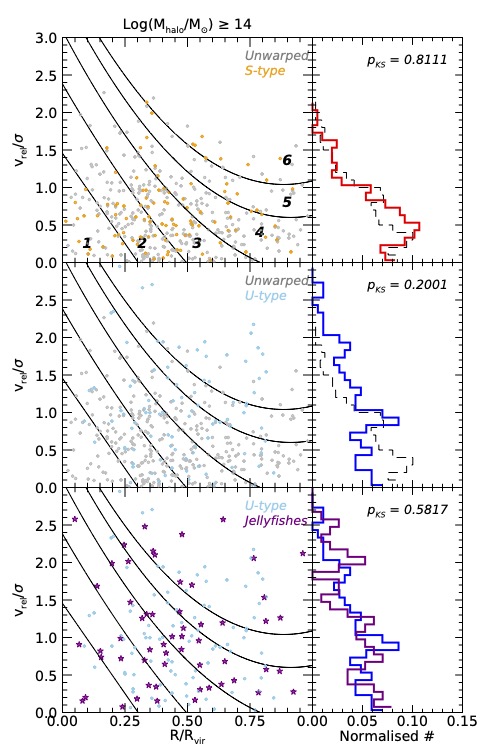}
    \caption{The same as Figure 11, but for galaxies in groups/clusters with massive host halo masses $M_{\textrm{halo}}/M_{\odot}$\,$\geq$\,$10^{14}$.}
    \label{fig:12}
\end{figure}

The spatial distribution and kinematics of galaxies in clusters allow us to trace the evolutionary phases of the infalling process.
This well-established method is based on the `phase-space diagram' (e.g., \citealt{2011MNRAS.416.2882M}; \citealt{2013MNRAS.431.2307O}; \citealt{2015MNRAS.448.1715J}; \citealt{2017ApJ...843..128R}; \citealt{2019ApJ...876..145S}; \citealt{2019MNRAS.484.1702P}).
The position where each galaxy is located on the diagram indicates its evolutionary phase from beginning of infall to a cluster to being entirely virialised. 
Figure~\ref{fig:11} shows galaxies with host halo mass of $M_{\textrm{halo}}/M_{\odot}$\,$\geq$\,$10^{12}$ on the phase-space diagram
along with the caustic regions (\citealt{2004A&A...414..445M}; \citealt{2011MNRAS.416.2882M}; \citealt{2019MNRAS.484.1702P}). 
Figure~\ref{fig:12} shows the same but for galaxies with more massive halo mass of $M_{\textrm{halo}}/M_{\odot}$\,$\geq$\,$10^{14}$.
We follow the definition of \citet{2019MNRAS.484.1702P} [see their equations (3) and (4) from p = 1 to 5]. 
The definition is valid for galaxies which are close to cluster center ($R/R_{\textrm{vir}} < 1.0$) and we use the subsample which are located in the same range. 
Region 1 is for virialised galaxies, whereas galaxies in Regions 5 and 6 begin to infall. 
As illustrated in Figure~\ref{fig:11} and Figure~\ref{fig:12}, while S-type warps are more concentrated in Region 1, the distribution of U-types is more extended to outer regions on the phase-space diagram.
For the looser criterion of halo mass, 22.4\% of S-types and 15.2\% of U-types are located in Region 1, respectively.
The difference becomes greater when we only use warped galaxies in more massive clusters; 35.1\% of S-type and 15.5\% of U-type warps are populated in Region 1.

Many studies showed that jellyfish galaxies tend to spread to the higher value of relative velocity on the phase-space diagram (e.g., \citealt{2018MNRAS.476.4753J}; \citealt{2019MNRAS.483.1042Y}). 
Intriguingly, S- and U-type warps among our sample show clearly different distribution of relative velocity. 
Histograms in Figure~\ref{fig:11} and Figure~\ref{fig:12} show that while S-type warps exhibit a similar distribution to unwarped control galaxies, U-type warps are more widely in relative velocity.
Among others, \citet{2019MNRAS.483.1042Y} classified $\sim$800 jellyfish galaxies from Illustris TNG100 simulation and showed that the relative velocity distribution of jellyfish galaxies are more extended than the undisturbed control sample. 
In the bottom panels of Figure~\ref{fig:11}, our U-type warps show an extended distribution in relative velocity similar to $\sim$\,150 jellyfish candidates observed by \citet{2016AJ....151...78P}. 
In the bottom panels of Figure~\ref{fig:12}, when we compare U-type warps and jellyfish candidates in more massive clusters, they still show similar distribution of relative velocity.
Also, we showed in Section~\ref{sec:Env} that, while the S-type warps show systematic correlations with the tidal interactions irrespective of cluster environments, U-types in clusters are not related with the local environments.
These results imply that most U-type warps in galaxy clusters are still not virialised within clusters' potential and seem to be affected by RPS like jellyfish galaxies.

\subsection{Stellar Mass, Star Formation Rate, and Gas Mass Fraction}

Jellyfish galaxies often exhibit increased star formation activity (\citealt{1985ApJ...294L..89G}; \citealt{2008MNRAS.388.1152P}; \citealt{2012MNRAS.427.1252M}; \citealt{2014MNRAS.438..444B}; \citealt{2016AJ....151...78P}; \citealt{2021arXiv210405383R}). 
The SF is enhanced at the infalling front of jellyfishes in clusters where gas compression occurs (\citealt{2009A&A...499...87K}; \citealt{2018ApJ...866L..25V}; \citealt{2019MNRAS.487.4580R}; \citealt{2021A&A...650A.111R}). 
For instance, \citet{2018ApJ...866L..25V} identified 42 RPS galaxies and found that both disks and tails have systematically higher SFR than the control sample. 
\citet{2020MNRAS.495..554R} investigated RPS galaxies in the Coma cluster to find that RPS galaxies exhibit a higher SFR relative to `normal' star-forming galaxies and isolated galaxies in fields. They suggested that RPS can trigger star formation prior to quenching.
\citet{2021arXiv211208728R} identified four jellyfish galaxies from the Perseus cluster and showed that all four jellyfishes exhibit star formation enhancement along the opposite direction of their stripped tails. 
\citet{2021gcf2.confE..22D} found 79 jellyfish candidates from the MACS0717 cluster and confirmed that jellyfishes tend to have higher sSFR and bluer colors. 
They stated that \textit{``jellyfish galaxy candidates appear to have somewhat larger SFRs than non-jellyfish star-forming galaxies."}
\citet{2022MNRAS.509.1342R} identified 48 RPS galaxies in low-mass groups ($M_{\textrm{halo}}/M_{\odot}$\,$<$\,$10^{14}$) and massive clusters ($M_{\textrm{halo}}/M_{\odot}$\,$\geq$\,$10^{14}$) by visual inspection using the Ultraviolet Near Infrared Optical Northern Survey (UNIONS) imaging and showed that RPS galaxies commonly have enhanced SFRs.

However, some studies claimed the lack of observational evidence of the SFR enhancement during RPS. 
For example, \citet{2021JKAS...54...17M} and \citet{2021IJAA...11...95H} respectively investigated 48 RPS galaxies in the Virgo cluster and 180 galaxies of the merging cluster Abell 3266, and found no strong evidence of RPS-induced global SF enhancement. 
\citet{2021IJAA...11...95H} suggested that RPS-induced SF enhancement is only locally modest, and the overall effect of SF quenching increases as the strength of RPS increases. 
However, the net effect of SF enhancement strongly depends on gas fraction and galaxies' evolutionary phase during RPS. 
Without morphological classification, these studies do not represent the characteristics of U-shaped disk galaxies in clusters.
\citet{2019MNRAS.487.3102G} introduced a case of jellyfish galaxy JO201 in which star formation is reduced during RPS. 
This galaxy shows a H$_2$ cavity with recently suppressed star formation by its AGN feedback in the last few $10^8$ yr. 
However, we exclude AGN-hosting galaxies in our sample in this study.

\begin{figure}
	\includegraphics[width=\columnwidth]{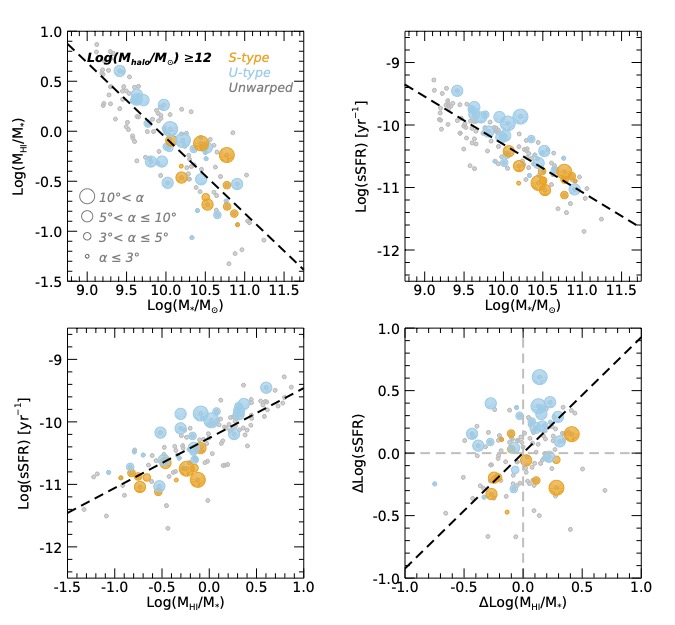}
    \caption{Distribution on the stellar mass-$f_{\textrm{HI}}$ (Upper, left panel), the stellar mass-sSFR (Upper, right panel), the $f_{\textrm{HI}}$-sSFR (Lower, left panel) and the $\Delta f_{\textrm{HI}}$-$\Delta$sSFR plane (Lower, right panel) for HI detected S-types (orange dots), U-types (cyan dots) and unwarped (grey dots) galaxies in groups/clusters with halo masses of $M_{\textrm{halo}}/M_{\odot}$\,$\geq$\,$10^{12}$.
    Symbol sizes represent the warping amplitudes as illustrated in the upper left panel. 
    Black dashed lines represent the linear fitted distribution of unwarped control galaxies.}
    \label{fig:13}
\end{figure}

\begin{figure}
	\includegraphics[width=\columnwidth]{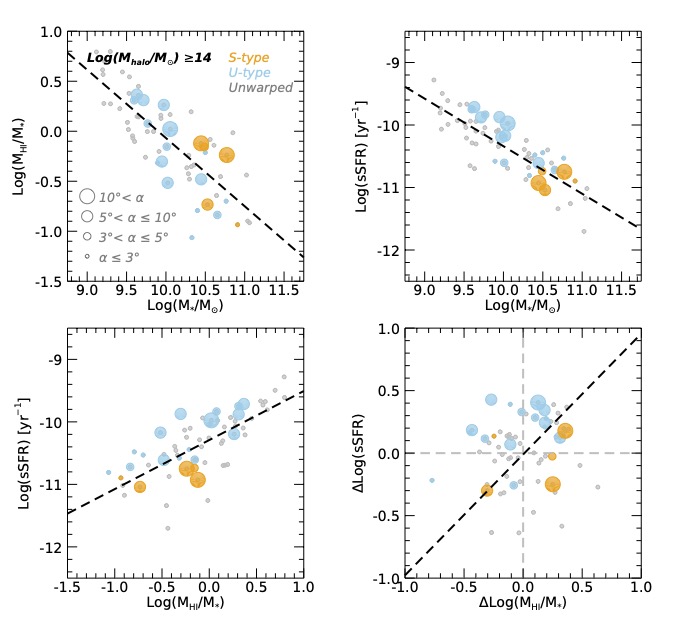}
    \caption{The same as Figure 13, but for galaxies in clusters with massive host halo masses of $M_{\textrm{halo}}/M_{\odot}$\,$\geq$\,$10^{14}$.}
    \label{fig:14}
\end{figure}

Now, we examine sSFR and star formation efficiency (SFE) of warped and unwarped galaxies at given stellar and gas mass.
Recent observations suggested that on average RPS galaxies exhibit lower stellar mass and higher gas fractions. 
For example, \citet{2018A&A...618A.130G} suggested that galaxies with lower stellar mass and higher gas fractions are more affected by RPS.
\citet{2021NatAs...5.1308G} found that $\sim$60\% of RPS galaxies show higher gas fractions. 
Even though infalling galaxies should lose their gas components during RPS, most stripped galaxies still exhibit higher gas fractions.
This result is consistent with theoretical expectations of RPS galaxies. 
\citet{2012MNRAS.420.1990S} demonstrated using their simulation that galaxies with higher gas mass fractions exhibit clearer signs of U-shaped stellar disk structures.
\citet{2019MNRAS.483.1042Y} found that jellyfishes are about three times more common at lower stellar mass ($M_{*}/M_{\odot} < 10^{10}$) than at higher mass ($M_{*}/M_{\odot} > 10^{10}$).
Following these recent findings, in the upper panels of Figure~\ref{fig:13} and Figure~\ref{fig:14}, we compare sSFRs and HI gas mass fractions ($f_{\textrm{HI}}=M_{\textrm{HI}}/M_{*}$) of galaxies in clusters, using different halo mass criteria ($M_{\textrm{halo}}/M_{\odot}$\,$\geq$\,$10^{12}$ and $M_{\textrm{halo}}/M_{\odot}$\,$\geq$\,$10^{14}$).
Our sample is matched with the galaxies in the ALFALFA survey (\citealt{2018ApJ...861...49H}) and S- and U-type warps have HI detection rates of $\sim$2.4\% and $\sim$3.6\%, respectively. 
Despite the small sample, U-type warps in clusters are, on average, of slightly smaller stellar mass and higher $f_{\textrm{HI}}$, which are susceptible to RPS.

The lower left panels of Figure~\ref{fig:13} and Figure~\ref{fig:14} show that, at the same $f_{\textrm{HI}}$, U-type warps exhibit higher sSFR than S-types and unwarped galaxies.
This is more significant for galaxies in more massive host halos ($M_{\textrm{halo}}/M_{\odot}$\,$\geq$\,$10^{14}$). 
The result indicates that U-type warps in clusters have high SFE similar to jellyfish galaxies (\citealt{2018ApJ...867L..29W}; \citealt{2019MNRAS.487.4580R}; \citealt{2019MNRAS.486L..26S}; \citealt{2020A&A...640A..22R}).
Using the EAGLE simulation, \citet{2016Galax...4...77T} found clear evidence of asymmetric SFE enhancement of RPS galaxies at the windward side.
\citet{2019MNRAS.486L..26S} simulated that pressure by intracluster medium can increase SFE under the assumption that galactic magnetic field halts evaporation of gas clouds during RPS. 
Observationally, \citet{2019MNRAS.487.4580R} and \citet{2020A&A...640A..22R} introduced the case of JO206 galaxy among the GAs Stripping Phenomena in galaxies with MUSE (GASP) sample which show higher SFE than field galaxies with similar stellar and gas mass. 
Specifically, RPS jellyfish galaxies exhibit 5--10 times higher SFE at disk than stripping tails due to compression of molecular gas by RPS.
Even at the high redshift universe, the enhancement of SFE of RPS galaxies is reported. 
\citet{2018ApJ...867L..29W} identified 14 member galaxies from a distant X-ray cluster CLJ1001 (z $\sim$ 2) and found a clear trend of their SFE as a function of cluster centric distance. 
Galaxies at the cluster center exhibit less molecular gas than field galaxies, but, intriguingly, higher SFR.
They suggested that cluster environment effects such as RPS can delay quenching with increasing SFE.

However, it is needed to consider that, in observations, one can detect not the initial but present $f_{\textrm{HI}}$. 
Many simulations suggested that RPS can remove the gas component efficiently, and RPS galaxies can change systematically from gas-rich to gas-poor over time. 
For example, \citet{2017ApJ...838...81Y} showed that RPS galaxies have a wide range of gas fractions. Using a phase-space diagram and HI morphologies, they defined the evolutionary phase of infalling galaxies and showed that the HI deficiency depends on the time since the first infall. 
Specifically, only recently infalled galaxies exhibit strongly disturbed HI morphologies and higher gas fractions. 
Similarly, \citet{2021JKAS...54...17M} showed that the HI deficiency of RPS galaxies in the Virgo cluster strongly depends on the evolutionary phase of infall.

\citet{2021arXiv211212244M} investigated RPS galaxies in the Coma cluster and showed using a simple model that the distribution of $\Delta f_{\textrm{HI}}$ and $\Delta$sSFR can trace the evolutionary phase of RPS. 
The residual of $f_{\textrm{HI}}$, $\Delta f_{\textrm{HI}}$, is defined by the difference in $f_{\textrm{HI}}$ between the warp sample and the mean $f_{\textrm{HI}}$ of its corresponding control sample. 
According to their model, gas-rich galaxies temporarily experience a short timescale ($\leq$ 300Myr) gas removal and starburst, resulting in the still regular HI gas fraction and enhanced SFR at the beginning of RPS. 
As galaxies evolve, the deficiency of HI increases and SFR is suppressed. 
Along this evolution during RPS, galaxies move from the first through second to third quadrant on the $\Delta f_{\textrm{HI}}$--$\Delta$sSFR parameter space.

The lower right panels of Figure~\ref{fig:13} and Figure~\ref{fig:14} show the distribution of S-, U-type warps and unwarped galaxies in clusters. 
In this $\Delta f_{\textrm{HI}}$--$\Delta$sSFR plane, the symbol size represents the warping amplitude.
Most U-types are distributed in the first and second quadrants, with only a few in the fourth quadrant.
The absence of galaxies in the fourth quadrant is similar to the RPS galaxies explored by \citet{2021arXiv211212244M}. 
We also find that more strongly warped U-types prefer the first quadrant.
On average, the mean value of warping amplitude of U-type warps in the first quadrant is $\sim$5.5$^{\circ}$ greater than that of U-types in the third quadrant regardless of the halo mass criteria of groups/clusters.
This is consistent with our aforementioned result on the correlation between sSFR enhancement and warping amplitudes for U-type warps.

\begin{figure}
	\includegraphics[width=\columnwidth]{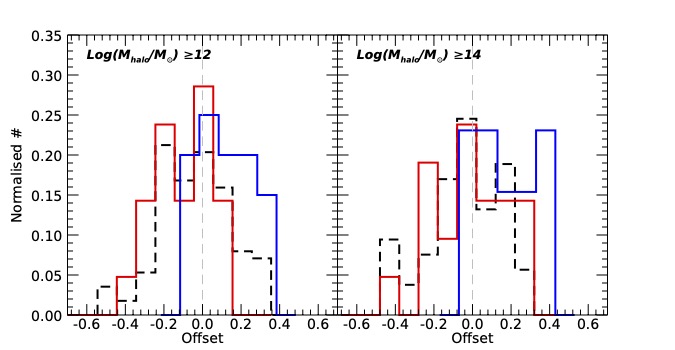}
    \caption{Distributions of the orthogonal $\Delta f_{\textrm{HI}}$-$\Delta$sSFR offsets of S-type (red), U-type warped (blue), and unwarped galaxies (black dashed) in groups/clusters with halo mass of $M_{\textrm{halo}}/M_{\odot}$\,$\geq$\,$10^{12}$ (left) and $M_{\textrm{halo}}/M_{\odot}$\,$\geq$\,$10^{14}$ (right).}
    \label{fig:15}
\end{figure}

To be more quantitative, we estimate the orthogonal offsets of S- and U-type warps from the unwarped galaxies. 
We fit the relation for unwarped galaxies in clusters with host halo mass of $M_{\textrm{halo}}/M_{\odot}$\,$\geq$\,$10^{12}$ [$\Delta \textrm{Log(sSFR)} = 0.97 \,\Delta\textrm{Log}(f_{\textrm{HI}}) - 0.01$] and in clusters with host halo mass of $M_{\textrm{halo}}/M_{\odot}$\,$\geq$\,$10^{14}$ [$\Delta \textrm{Log(sSFR)} = 0.94\,\Delta\textrm{Log}(f_{\textrm{HI}})$].
The resulting distributions of offsets are illustrated in Figure~\ref{fig:15}.
On average, U-type warped galaxies show $\sim$ 0.2 dex higer offset than unwarped and S-type ones. 
This result is consistent with theoretical expectation of \citet{2021arXiv211212244M}.

\section{Summary and Discussion}
\label{sec:conc}

Our main questions in this study are: ($a$) are S- and U-type warps created by the same mechanism? and ($b$) how can we explain U-type warps? 
To address the questions, 
we complete the most extensive catalog of $\sim$3000 nearby (0.02\,$<$\,z\,$<$\,0.06) massive ($M_{*}/M_{\odot}$\,$>$\,$10^9$) warped disk galaxies through our new automatic warp measurement scheme.
Then, we compare key properties, including optical colors, sSFR, several environmental parameters, warping amplitudes, and kinematics within groups/clusters, stellar mass and gas fraction, between S- and U-type warped galaxies for the first time.

Our findings are summarised as follows.
\begin{enumerate}
\item
U-type warps exhibit bluer optical color and higher sSFR than S-types and unwarped galaxies at a given stellar mass.
The $\Delta$Log(sSFR) correlates positively with warping amplitudes.
The results indicates that U-type warp formation mechanism entails higher sSFR.
\item
While warp properties of S-type warps are correlated with the tidal force by the nearest neighbors irrespective of galaxy cluster membership, U-types in clusters show no local environmental dependence. 
This implies that the conventional gravitational interactions create S-type warps only and other non-tidal alternative mechanisms are required to explain the existence of U-type warps at least in clusters.
\item 
A thorough visual inspection of jellyfish galaxies in literature leads us to find some intriguing examples of RPS-driven U-shaped warped galaxies bent into the opposite direction of stripped gas tails, which is consistent with previously published theoretical expectations of RPS-driven stellar warps.

\item 
There are considerable similarities between U-type warps in galaxy clusters and RPS-induced jellyfish galaxies in terms of the morphology, location on the phase-space diagram, sSFR and $f_\textrm{{HI}}$.
The results suggest that U-types in groups/clusters could be connected to jellyfish galaxies at the very early stage of RPS, explaining the existence of U-types which are hard-to-make in conventional galaxy--galaxy interaction simulations.
\end{enumerate}

The discovery of similarities between U-type warps in groups/clusters and jellyfish galaxies is encouraging in the context of unveiling the origin of warped disk galaxies.
Given that only 17.5$\%$ (4.1$\%$) of U-type warps belong to groups/clusters with $M_{\textrm{halo}}/M_{\odot}$\,$\geq$\,$10^{12}$ (more massive clusters with $M_{\textrm{halo}}/M_{\odot}$\,$\geq$\,$10^{14}$), the RPS origin of the group/cluster U-type warps remains to be approved by follow-up observations such as integral field units and HI survey of the gas component in these galaxies.
Also, it is still limited to explain the other majority of U-type warps which live in fields.
Different pathways to form U-shaped disk galaxies including galaxy-galaxy interactions, large-scale gas infall, and interactions with misaligned dark matter halos remain to be fully explored.

We propose that RPS galaxies can be observed as U-type warps with higher sSFR and stronger warping amplitudes at the beginning of infall. 
Following our scenario, it is natural to expect that U-type warps should be more common in more massive clusters due to their stronger effect of RPS.
However, the incidence and strength of U-type warps in groups/clusters do not depend on their host halo mass. 
The absence of a strong correlation between U-type warps and groups/clusters' halo mass can be explained by the short time-scale of RPS-driven U-shaped stellar disks.
Theoretical studies expected that U-shaped stellar disks could be constructed very briefly ($\leq$ 200Myr) at the early stage of RPS (\citealt{2012MNRAS.420.1990S} and \citealt{2022arXiv220101316L}).
This time-scale is too short to be observed as stable U-shaped stellar disks. 
In contrast, RPS occurs slowly with a longer time-scale ($\sim$3 Gyr) in galaxy groups with less massive halos.
Thus, despite its lower efficiency of RPS, there is more chance to detect stable stripped stellar disks in less massive groups than in more massive clusters.
It is necessary to investigate the time-dependence evolution of RPS-driven stellar warps in detail by further cosmological simulations.

Our results are essential for galactic warp studies because U-shaped RPS galaxies can lead to overestimation of the incidence of tidal-origin U-type warps.
Thus, warped galaxies in fields and clusters should be investigated separately to assess the effect of the tidal interactions without the contamination of cluster environment effects.
Also, it is important for RPS studies that suggest to look for warps as an additional sign of stripping.

The measurement of warp morphologies and warping amplitudes are affected by the complex spiral arm, dust lanes, and the orientation angle from the observer's viewpoint.
We thus need more massive data to study the effect of the substructures on warp measurements. 
We initiated \textit{Poppin' Galaxy}\footnote{\href {https://www.zooniverse.org/projects/wim0705/poppin-galaxy}{https://www.zooniverse.org/projects/wim0705/poppin-galaxy}} project through the Zooniverse in 2018 to gather massive warped galaxies classified by over 10,000 volunteers.
In forthcoming papers, we are to investigate the origin and properties of the warp phenomenon exploiting our extensive observational data in comparison to high-resolution cosmological simulations of galaxies.

\acknowledgments
S.-J.Y. acknowledges support by the Mid-career Researcher Program (No. 2019R1A2C3006242) through the National Research Foundation of Korea.

Funding for the Sloan Digital Sky Survey (SDSS) and SDSS-II has been provided by the Alfred P. Sloan Foundation, the Participating Institutions, the National Science Foundation, the U.S. Department of Energy, the National Aeronautics and Space Administration, the Japanese Monbukagakusho, the Max Planck Society, and the Higher Education Funding Council for England. The SDSS Web site is \url{http://www.sdss.org/}.

The SDSS is managed by the Astrophysical Research Consortium (ARC) for the Participating Institutions. The Participating Institutions are the American Museum of Natural History, Astrophysical Institute Potsdam, University of Basel, University of Cambridge, Case Western Reserve University, The University of Chicago, Drexel University, Fermilab, the Institute for Advanced Study, the Japan Participation Group, The Johns Hopkins University, the Joint Institute for Nuclear Astrophysics, the Kavli Institute for Particle Astrophysics and Cosmology, the Korean Scientist Group, the Chinese Academy of Sciences (LAMOST), Los Alamos National Laboratory, the Max-Planck-Institute for Astronomy (MPIA), the Max-Planck-Institute for Astrophysics (MPA), New Mexico State University, Ohio State University, University of Pittsburgh, University of Portsmouth, Princeton University, the United States Naval Observatory, and the University of Washington.

\bibliography{sample63}
\bibliographystyle{aasjournal}


\begin{table}
	\centering
	\caption{Basic statistics of warped disk galaxies in our sample.}
	\label{tab:example_1}
	\begin{tabular}{lcccr} 
		\hline
		Morphology & Incidence & Number Fraction & Median Warp Angle & Median Log($M_{*}/M_{\odot}$) \\
		\hline
		S-type & 2206 & 27.6\% & $5.45^{\circ} \pm 0.07^{\circ}$ & 10.39 $\pm$ 0.42 \\
		U-type & 1456 & 18.2\% & $4.37^{\circ} \pm 0.08^{\circ}$ & 10.36 $\pm$ 0.43 \\
		\hline
	\end{tabular}
\end{table}

\clearpage

\clearpage

\clearpage

\clearpage

\clearpage

\clearpage

\clearpage

\end{document}